\documentclass[manuscript]{acmart}

\usepackage{balance}
\usepackage{caption}
\usepackage{subcaption}
\usepackage{mwe, chngpage, booktabs}

\AtBeginDocument{%
  \providecommand\BibTeX{{%
    \normalfont B\kern-0.5em{\scshape i\kern-0.25em b}\kern-0.8em\TeX}}}

\copyrightyear{2021}
\acmYear{2021}
\setcopyright{acmlicensed}\acmConference[MM '21]{Proceedings of the 29th ACM International Conference on Multimedia}{October 20--24, 2021}{Virtual Event, China}
\acmBooktitle{Proceedings of the 29th ACM International Conference on Multimedia (MM '21), October 20--24, 2021, Virtual Event, China}
\acmPrice{15.00}
\acmDOI{10.1145/3474085.3475631}
\acmISBN{978-1-4503-8651-7/21/10}

\usepackage{color,soul}

\newcommand{\tildeapprox}{\raise.17ex\hbox{$\scriptstyle\mathtt{\sim}$}}
\usepackage{tabularx}

\begin{document}
\fancyhead{}

\title{Using Interaction Data to Predict Engagement with Interactive Media}

\author{Jonathan Carlton}
\orcid{1234-5678-9012}
\email{jonathan.carlton@manchester.ac.uk}
\affiliation{%
    \institution{Dept. of Computer Science, University of Manchester \& BBC R\&D, Manchester, UK}
}

\author{Andy Brown}
\email{andy.brown01@bbc.co.uk}
\affiliation{%
    \institution{BBC R\&D, Manchester, UK}
}

\author{Caroline Jay}
\email{caroline.jay@manchester.ac.uk}
\affiliation{
    \institution{Dept. of Computer Science, University of Manchester, UK}
}

\author{John Keane}
\email{john.keane@manchester.ac.uk}
\affiliation{
    \institution{Dept. of Computer Science, University of Manchester, UK}
}

\begin{abstract}
Media is evolving from traditional linear narratives to personalised experiences, where control over information (or how it is presented) is given to individual audience members. Measuring and understanding audience engagement with this media is important in at least two ways: (1) a post-hoc understanding of how engaged audiences are with the content will help production teams learn from experience and improve future productions; (2), this type of media has potential for real-time measures of engagement to be used to enhance the user experience by adapting content on-the-fly. Engagement is typically measured by asking samples of users to self-report, which is time consuming and expensive.  In some domains, however, interaction data have been used to infer engagement. Fortuitously, the nature of interactive media facilitates a much richer set of interaction data than traditional media; our research aims to understand if these data can be used to infer audience engagement. In this paper, we report a study using data captured from audience interactions with an interactive TV show to model and predict engagement. We find that temporal metrics, including overall time spent on the experience and the interval between events, are predictive of engagement. The results demonstrate that interaction data can be used to infer users' engagement during and after an experience, and the proposed techniques are relevant to better understand audience preference and responses.
\end{abstract}




\maketitle

\section{Introduction}\label{sec:introduction}
Interactive media experiences present individual audience members with media that is tailored to their particular context, knowledge, or needs. It is different from traditional media - where a single version of the content aims to engage all members of the audience - by varying either the information presented or the form of its presentation.  The information used to determine the variation can come from either (1) users interacting with the content as they consume it; (2) collecting it from the user at the start of the experience (thereby allowing a more `lean-back' mode of consumption); (3) an existing `user profile', (4) by a combination. In this paper the focus is on the first of these. The interactive nature of these experiences makes them more complex than traditional forms of media.  It is therefore important for production teams to monitor how successful or otherwise they have been, so that they can identify and rectify issues with the content, and learn how to improve the user experience for future productions.

A key metric for success is how engaged the audience members were with the experience.  However, user engagement is a complex, multifaceted phenomenon \cite{attfield2011towards} and as media is typically consumed by a geographically distributed audience, in different contexts and on different devices, accurate measurement has both practical and methodological challenges.  Interaction data have been used in some domains as an unobtrusive and scalable way of monitoring engagement \cite{attfield2011towards,lehmann2012models,doherty2018engagement}; our research is directed at understanding if it can also be used in this domain. Given that interactive media experiences involve greater interactivity with the content than traditional media, is it possible to monitor these interactions and use them to infer engagement? The study explores and presents new methods of collecting information and analysis techniques that broadcasters could use to better understand their audiences and how engaged they were with their content; we are not evaluating an experience, we explore whether interaction data might be useful when measuring engagement.

An additional motivation for being able to measure engagement through interaction data is that this could allow near real-time monitoring of engagement. Thus, not only would it be possible to perform retrospective analysis to improve future work, it could also enable the content to adapt on-the-fly. For example, audience members who appear highly engaged may be offered supplementary material, while those who appear to have low-levels of engagement might be given a shorter or simpler conclusion to the content. Designing successful interventions of this sort is clearly demanding, but could significantly enhance the user-experience and is only really possible if engagement can be monitored real-time.

Interaction data have been used successfully in other domains, namely online news \cite{lagun2016understanding,grinberg2018identifying,lu2018between,constantinides2018framework} and search \cite{diriye2012leaving,kim2014modeling,zhuang2017understanding}. It has also been used to extract user behaviours as proxies for engagement \cite{barbieri2016improving,arapakis2016predicting} and satisfaction \cite{chuklin2016incorporating,mehrotra2017user}. As techniques and measures for engagement are untested in this domain, we study a production-quality and nationally released interactive TV show where the audience are given control over the presentation and their path through the story. We collected ground truth engagement metrics and user interactions (code and data are available: \cite{jonathan_carlton_2021_5137806}) from the live and in-the-wild interactive TV show to address the question:

\begin{itemize}
    \item[\textit{(RQ)}] What behaviours can we infer from interaction metrics that help in understanding how engaged an audience member is with an interactive media experience?
\end{itemize}

We extracted a range of metrics and investigated how these relate to the engagement levels reported by audience members. To understand the relationships between metrics and engagement levels, we trained an interpretable model to predict engagement and derive feature importance using internal weights and a model-agnostic interpretation technique. We found that temporal metrics -- the time between events and the time an audience member takes to complete the story -- effectively distinguishes, and are strongly associated with, engagement levels. Our findings demonstrate the utility of interaction data for offering media creators insights into audience engagement and that it can be directly inferred from the data alone. Overall, we make the following contributions:

\begin{itemize}
    \item We confirm a link between user engagement and interaction metrics collected from an interactive media experience;
    \item We demonstrate the possibility of predicting engagement levels from the interactions of users;
    \item We establish, more generally, the utility of collecting and analysing user interactions from interactive media experiences and the potential insights about user engagement that can be garnered to feed into the creative process.
\end{itemize}

The remainder of the paper is structured as follows: Section \ref{sec:background} discusses related work. Section \ref{sec:methods} describes the study, and Section \ref{sec:results} presents the results. Section \ref{sec:discussion} discusses and contextualises the findings and Section \ref{sec:conclusion} draws conclusions and considers future work.
\section{Related Work}\label{sec:background}
We aim to find approaches that content creators can use to understand how audiences respond to their content. We hypothesise that interaction with interactive media can be used to infer certain types of audience behaviour, and such behaviours can in turn be related to how engaged an audience member felt. The literature review, therefore, is organised into three sections: (1) the various approaches to measuring engagement; (2)  interaction data - what has been collected and how it has been used; (3) the link between the two - how interaction has been used to infer engagement. 

\subsection{User Engagement}
User Engagement (UE) is a complex, multifaceted phenomenon \cite{attfield2011towards} that can be broadly defined as a quality of user experience characterised by the depth of a user's investment when interacting with a digital system \cite{o2016theoretical}. Capturing a reliable ground truth metric for engagement often requires directly questioning the audience, while physiological approaches (electrocardiograph \cite{arapakis2017interest,belle2011physiological} and eye-tracking \cite{nakano2010estimating}) show potential as effective proxies \cite{doherty2018engagement}. A popular and frequently used survey is the User Engagement Scale (UES) \cite{o2010development,o2018practical} which measures engagement with digital technology through a combination of four factors: focused attention, perceived usability, aesthetic appeal, and reward. An alternative to UES which can provide a more detailed picture is the Engagement Sampling Questionnaire \cite{schoenau2011hooked}, where participants are regularly sampled during a study to capture how engagement changes over time. 

\subsection{Interaction Data}
Self-reported engagement measures, whilst reliable can become impractical to use at scale and in-the-wild. Interaction data, in comparison, is straightforward to capture from many users \cite{attfield2011towards,lehmann2012models}, describes user behaviour without disrupting the experience, and removes the burden of retrospectively self-reporting engagement \cite{attfield2011towards}. When interaction data is collected, it typically consists of fine-grained interactions, such as mouse cursor events \cite{arapakis2017interest,lagun2016understanding,zhuang2017understanding} or mobile interactions (for example, swipe direction) \cite{constantinides2018framework,lu2019quality}. However, as engagement is measured indirectly, there is greater potential for error in inference; further, when users are aware interactions are being collected, behaviour may change \cite{doherty2018engagement}. In this work, we collect interaction data that take the form of interface-level actions, for example clicking the play/pause button or changing window orientation (see Section \ref{sec:data-collection}). Typically, interaction metrics are extracted from raw interaction data and take the form of descriptive statistics, such as the number of page visits, dwell time, or the number of clicks \cite{hong2019tutorial}. From these metrics, users' behaviours can be inferred and used as proxies for the phenomenon of interest \cite{kim2014modeling,barbieri2016improving,miroglio2018effect}.

\subsection{From Data to Phenomenon}
As measuring engagement with interactive media experiences is a new area, here we present and learn from approaches used in domains where understanding user behaviour through their interactions is more established, examining: the types of data collected; the behaviours inferred; and what analysis techniques were applied.

\subsubsection{Online News}
Online news reading is a related area, where personalising and tailoring the reading experience to different users` preference is important. In \cite{lagun2016understanding}, sequences of reading patterns were reconstructed from scrolling events and temporal statistics to provide a more nuanced approach to understanding user attention beyond dwell time. Common reading behaviours were identified and significant differences found with respect to interaction metrics; for example, users determined to be in the completed group spent significant time in the comment section and recorded an increase in the number of clicks on an article compared to the other three detected behaviours (bounce, shallow, and deep). In a similar approach, \cite{grinberg2018identifying} extracted dwell time, maximal reading depth, active engagement, the proportion of an article visible on screen, scrolling speed, and normalised engagement; from this they built a probabilistic model to identify and predict six article reading behaviours: bounce back, shallow, scan, idle, read, and read long. In both these works, behaviours were claimed to be directly mapped to engagement, with the former concentrating on focused attention - the ability of the content to capture the attention of readers, a dimension of engagement \cite{o2010development,o2018practical} that is measured by UES. However, neither verified that their behaviours were representative of a ground truth measure of engagement and instead demonstrated the model's ability to predict reading behaviours accurately. Similarly, \cite{arapakis2014understanding} identified that certain types of mouse gesture clusters negatively correlated with focused attention, but did not reveal what combination of interaction metrics determined the clusters, limiting insight into the specific behaviours linked to attention.

In contrast, both \cite{arapakis2017interest} and \cite{lu2018between} captured ground truth measures of a phenomenon from news article readers and explored differences in metrics to uncover user behaviours indicative of the phenomenon in question. The former \cite{arapakis2017interest} collected ECG data along with focused attention from UES and found links between news articles, attention, and cognitive behaviours such as motivation to approach or escape (derived from the ECG data). The latter, \cite{lu2018between}, extracted temporal, reading and scrolling metrics from user interactions with articles. By applying correlation analysis between user groups (determined by the ground truth measure), the authors found users who liked an article tended to read more, at a slower pace, and revisited content.

\subsubsection{Search}
The search and information retrieval domains both use the interactions of users to understand, model, and predict phenomena. Often user satisfaction, analogous to user engagement in a search context \cite{hong2019tutorial}, is the focus. Using the novelty sub-scale of UES, \cite{zhuang2018can} investigate differences between sequences of interactions and find that a user clicking on the next search result page followed by a click on a result can effectively discriminate between high and low novelty. Similarly, in exploring whether interaction sequences can be predictive of satisfaction, \cite{mehrotra2017user} found that a click followed by a long dwell time is correlated with satisfaction; in contrast, moving around the search engine result page is associated with dissatisfaction. In \cite{zhuang2017understanding}, UES sub-scales were used as separate prediction targets to investigate behaviour during a search task. Interaction metrics describing sessions were extracted and grouped into categories: click, query, result, and time. Through calculating how much a metric contributes to the prediction of each sub-scale, the authors found that the amount of time spent searching on a results page is related to the usability sub-scale and that metrics related to the query and time are predictive across all four sub-scales.

Abandonment, where a user does not click any search results \cite{li2009good}, is another type of phenomenon of interest. For example, {\it good} (the user finds what they are looking for) and {\it bad} (they do not) abandonment are explored in \cite{williams2017does}. The authors found, by comparing interaction metrics between two types of abandonment, that an increase in scrolling, longer dwell times, and other events were associated with bad abandonment. The rationale behind a user abandoning a search is explored by \cite{diriye2012leaving} in a retrospective survey used to find common reasons for abandoning. They found that the amount of time taken by a user to interact with search results and the quality of the result are strong predictors of dissatisfaction and subsequent abandonment; a similar result is found by \cite{song2014context}, with longer session lengths being associated with bad abandonment. 
\section{Methodology}\label{sec:methods}
\begin{figure*}[h]
    \centering
    \begin{subfigure}[b]{0.49\textwidth}
        \includegraphics[width=\textwidth]{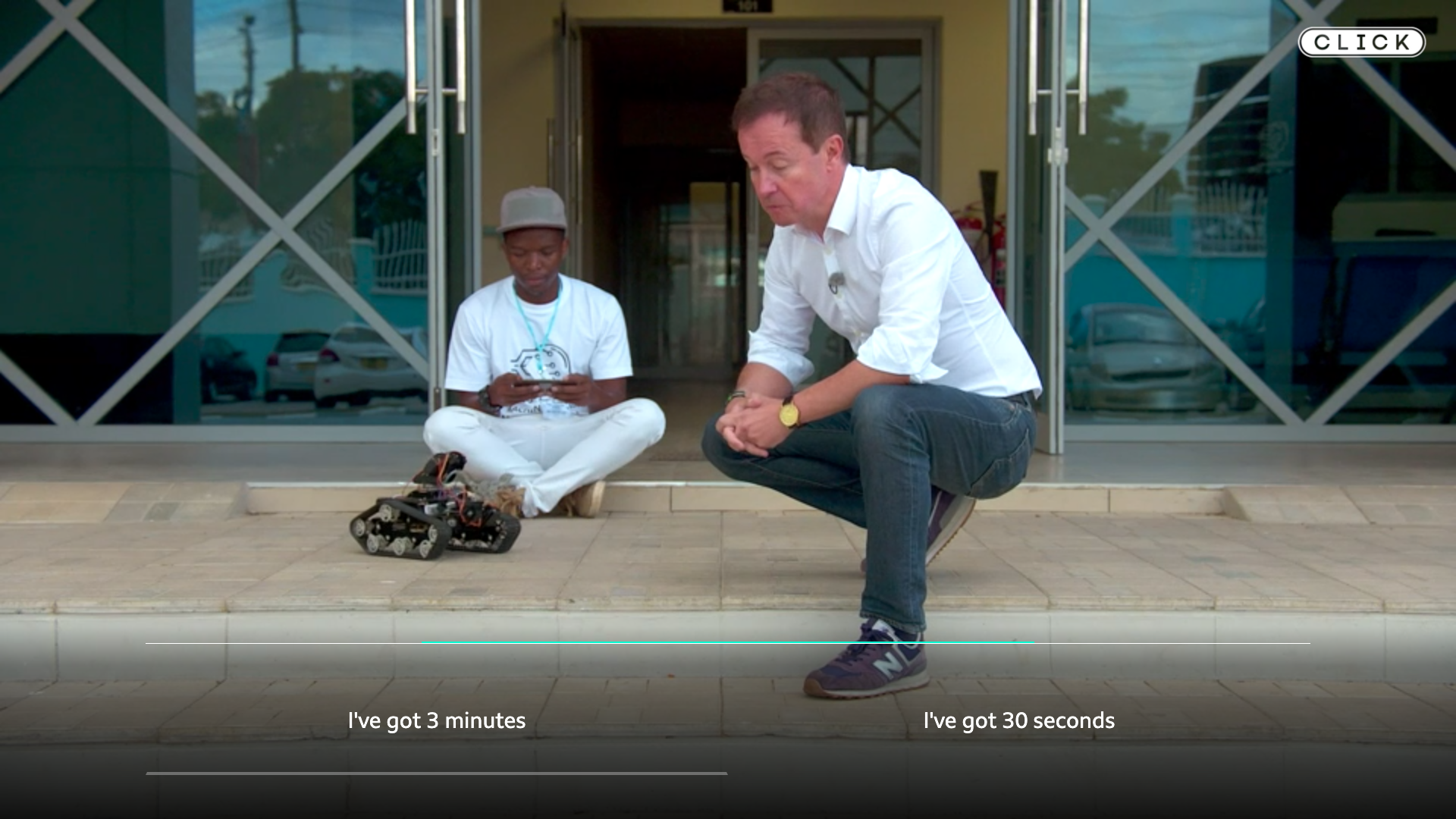}
        \Description{There is a make a choice graphic on screen and the audience is asked to 
        pick between whether they are ready or if they want to do this}
    \end{subfigure}
    \hfill 
    \begin{subfigure}[b]{0.49\textwidth}
        \includegraphics[width=\textwidth]{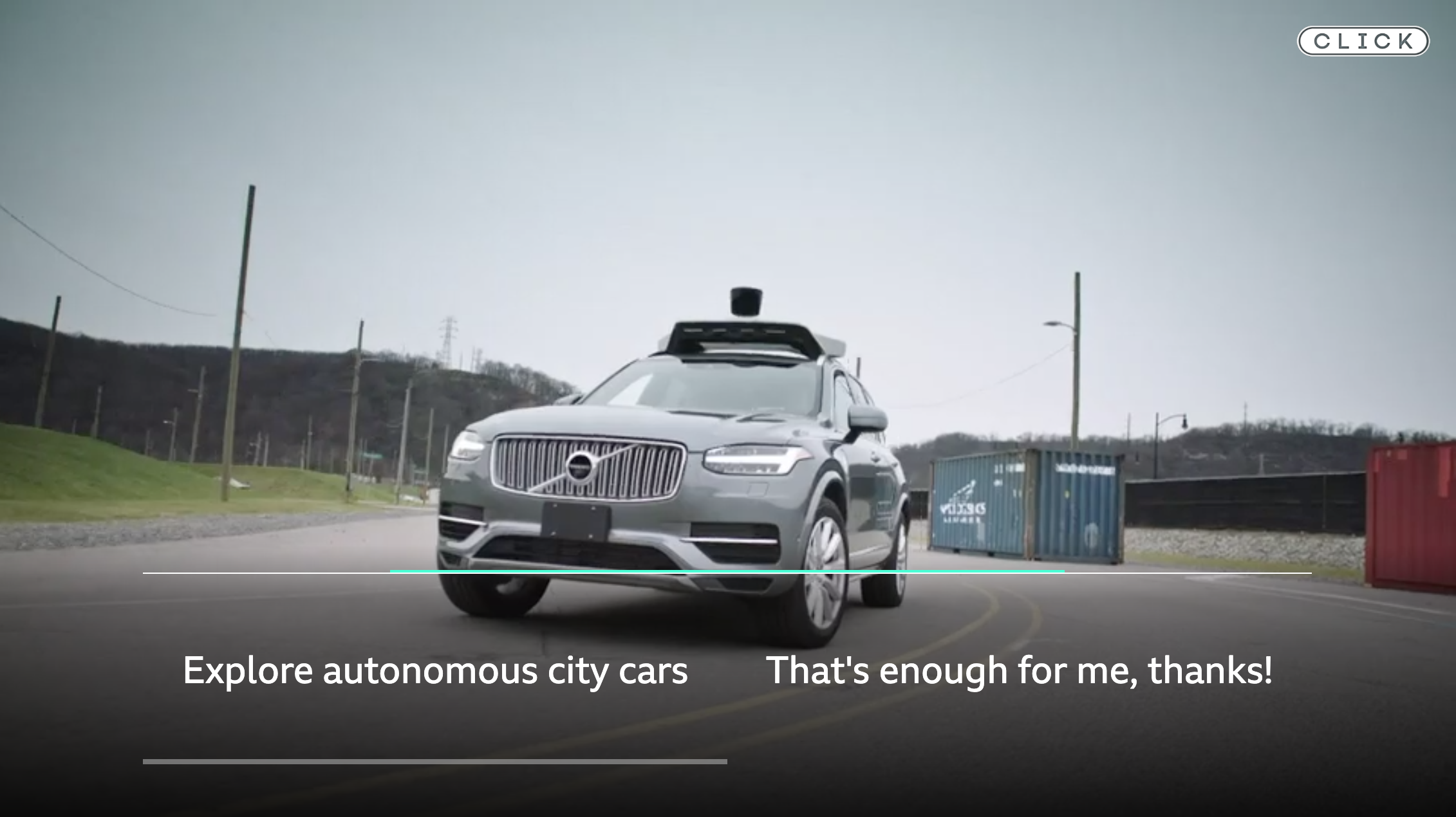}
        \Description{An autonomous car is on the screen and the audience can choose either 
        exploring autonomous city cars or if they have had enough of the show segment}
    \end{subfigure}
    \caption{Examples of choices given to the audience in the interactive episode of Click: on the left, viewers can choose between a long or a short version, while on the right they can either explore another aspect of a topic or move on.}
    \label{fig:click-overview}
\end{figure*}

In this section, we describe the study carried out on an interactive TV show called Click, which was a one-thousandth special edition of a long-running BBC technology show in the UK. This particular episode of Click was an interactive form of storytelling (known as an interactive branching narrative), where the user determines their path through the story based on their interests. The show is formed of four substories covering different topics; the two main ones are about technology use in Malawi and autonomous vehicles. The audience can control what they see both between and within these substories, e.g., you can choose to view one or both use-cases for the autonomous vehicle technology (industrial and/or consumer) as well as having the option to go into more or less detail about the technology. The decisions that the audience are asked to make are prompted by the on-screen host (shown in Figure \ref{fig:click-overview}). If the audience decides not to interact, then a default path is automatically followed. Behind the scenes, the show is broken down into narrative elements - with a single narrative element containing a short scene in the show - which when brought together form the entire show. The audience has control over the show and video content, such that they can navigate between narrative elements using the next and back buttons and replaying or rewinding the current narrative element. Standard video controls are also available: play/pause, full-screen, volume control, and video scrubbing. All of the controls trigger events which are logged by built-in analytics; for example, when the user moves from one narrative element to another is logged as a narrative element change event (detailed in Section \ref{sec:data-collection}).

\subsection{Study Design}
To understand the relationship between user behaviour and engagement with an interactive branching narrative, we designed a study to collect both interaction data and ground truth engagement metrics. This study was designed to be opt-in and in-the-wild, set in a live production environment attached to the official national release of Click. Interaction data were collected anonymously for all users of the experience (for which consent was given). In addition, after the closing credits, viewers were given the opportunity to complete a short survey that measured their engagement with the content; this was anonymous and did not collect any personal or demographic data, but could be associated with the interaction data.  This approach allowed us to collect data from a relatively large number of users in an ecologically valid setting and respecting privacy; the downside was that the users for whom engagement data were available was a self-selecting group from those who had reached the end of the show.

\subsection{Data Collection}\label{sec:data-collection}
We collected interaction data using built-in analytics, which log when a user performed an interface action or when a contextual change occurred (a window orientation or browser visibility change). The data takes the following form: \textit{user\_id} - an anonymous identification string, \textit{timestamp} - millisecond granularity timestamp, \textit{action\_type} - the type of event that occurred, \textit{action\_name} - the button clicked/context change, and \textit{data} - additional metadata about the event, for instance hidden/visible for browser visibility changes. The analytics capture the following events: play/pause, back, next, fullscreen, subtitles, volume, video scrub, seek backwards, seek forwards, browser visibility change, window orientation change, narrative element change, and link choice.

To capture engagement metrics, we used the UES survey due its wide usage and validation in a range of contexts \cite{doherty2018engagement}. As the survey was attached to a live production system, we used the short-form 12 questions version, rather than the 30 questions. The survey was administered post-credits, to ensure it did not disrupt the experience. Following the guidance in \cite{o2018practical}, we altered the questions to fit with the Click experience, for example, changing ``Using Application X was worthwhile'' to ``Using this interactive episode of Click was worthwhile''. We followed the UES guidance \cite{o2018practical} to calculate engagement scores: four individual factor means were calculated, followed by an overall mean to obtain a final score.

\subsection{Interaction Metrics}
As collected interaction data tells us little about user behaviours, we aggregated the data into statistical interaction metrics. We extracted 57 metrics per user which included individual event counts, relative frequencies of events, and the total number of user actions. We also derived temporal statistics: the time to completion (amount of time it took the audience member to reach a defined endpoint in minutes), session length (the overall time spent on Click in minutes), and hidden time (the time spent with the browser window hidden). The types of pauses between events were also captured: short (between one and five seconds), medium (between six and 15), long (between 16 and 30), and very long (more than 30), as inspired by \cite{mehrotra2017user,williams2017does} and using the values defined in \cite{williams2017does}. In calculating session length and time to completion, we subtracted the hidden time as we were interested in the amount of active time that the user spent on the experience and in reducing the number of outliers caused by large session lengths. For the time to completion metric, we chose a node in the Click experience story graph where all sub-stories are brought back together, situated just before the final scenes of the show. Our justification for choosing this endpoint is that it is where all paths through the story graph meet and the main portion of the narrative finishes, with only the credits remaining.

\subsection{Analysis}

\subsubsection{Discovering Relationships}
To find relationships and differences between interaction metrics and levels of engagement, we performed correlation analysis, tested for statistical differences between levels of engagement reported by users, and investigated metrics that discriminate between high and low engagement.

To determine which correlation test to perform, we computed Shapiro-Wilk`s test for normality. For non-parametric metrics, we calculated Spearman`s rank-order correlation coefficient ($r_s$), while for normally distributed metrics, we used Pearson`s $r$ correlation coefficient. We performed the Mann-Whitney $U$ test and common language effect size to test for differences between metrics and engagement levels, corrected for multiple tests using False-Discovery Rate. Engagement levels were determined by splitting users on their median UES scores \cite{o2013mixed} creating high and low groups.

Following a similar approach to \cite{mehrotra2017user,zhuang2018can}, we applied the chi-square test ($\chi^2$) to compute the discriminatory power of the interaction metrics between high and low engagement. The chi-square test evaluates the probability of the observed result given the null hypothesis being true. As such, we propose the following null ({\bf H0}) and alternative ({\bf HA}) hypothesises:

\begin{itemize}
    \item[{\bf H0}] There are no detectable differences in the interaction metrics between the 
    two engagement classes (low and high).
    \item[{\bf HA}] There exist detectable differences in the interaction metrics between the 
    two engagement classes (low and high). 
\end{itemize}

Findings were considered significant at $p < .05$ and calculated using Python 3.6 with the SciPy \cite{2020SciPy-NMeth} and Pandas \cite{mckinney2010data} libraries.

\subsubsection{Predicting Engagement}
To investigate if latent relationships in the interaction metrics are predictive of engagement, we trained interpretable models to predict high or low engagement, extracted feature importance, and applied a model-agnostic interpretability method called Shapely Additive Explanations (SHAP) \cite{lundberg2017unified}. To account for the differences in ranges between metrics, e.g. session length (minutes) and relative frequencies (zero to one), we scaled each sample to a range between zero and one. Interaction metrics that contained more than 50\% zeros were converted into binary.

Using the processed interaction metrics, we evaluated a range of interpretable models using 10-fold stratified cross-validation with Area Under the Curve (AUC) as the performance metric. We chose AUC as we wish to evaluate the model`s ability to distinguish between two classes. The algorithms evaluated were: Logistic Regression, Gaussian Na\"ive Bayes, Decision Tree, Linear Discriminant Analysis, K-Nearest Neighbours, Support Vector Machine (SVM), and an SVM with stochastic gradient descent training. These algorithms were chosen as they are primarily interpretable, each make different assumptions about the data, and are proven to work well on a range of different prediction tasks. We did not use more complex models, such as Neural Networks, due to added complexity in terms of interpretability and structure. Once the best performing model was identified, we then performed a 10-fold stratified cross-validated grid search to find the optimal hyper-parameters and trained an optimised model using a standard train-test split ($80:20$) to fit the model used to form the investigation. 

To evaluate the importance of interaction metrics in predicting high/low engagement, we extracted feature weights from the model and calculated SHAP values, which provide a model-agnostic measure of how a feature value impacts the model prediction and uses a weighted linear regression (for further details on SHAP, see \cite{lundberg2017unified}). Pre-processing and modelling were performed using Python 3.6 with the scikit-learn \cite{scikit-learn} and the SHAP libraries \cite{lundberg2017unified}.

\section{Results}\label{sec:results}

\subsection{Sample}
\begin{figure}
    \centering
    \begin{subfigure}[b]{0.49\textwidth}
        \centering
        \includegraphics[width=\textwidth]{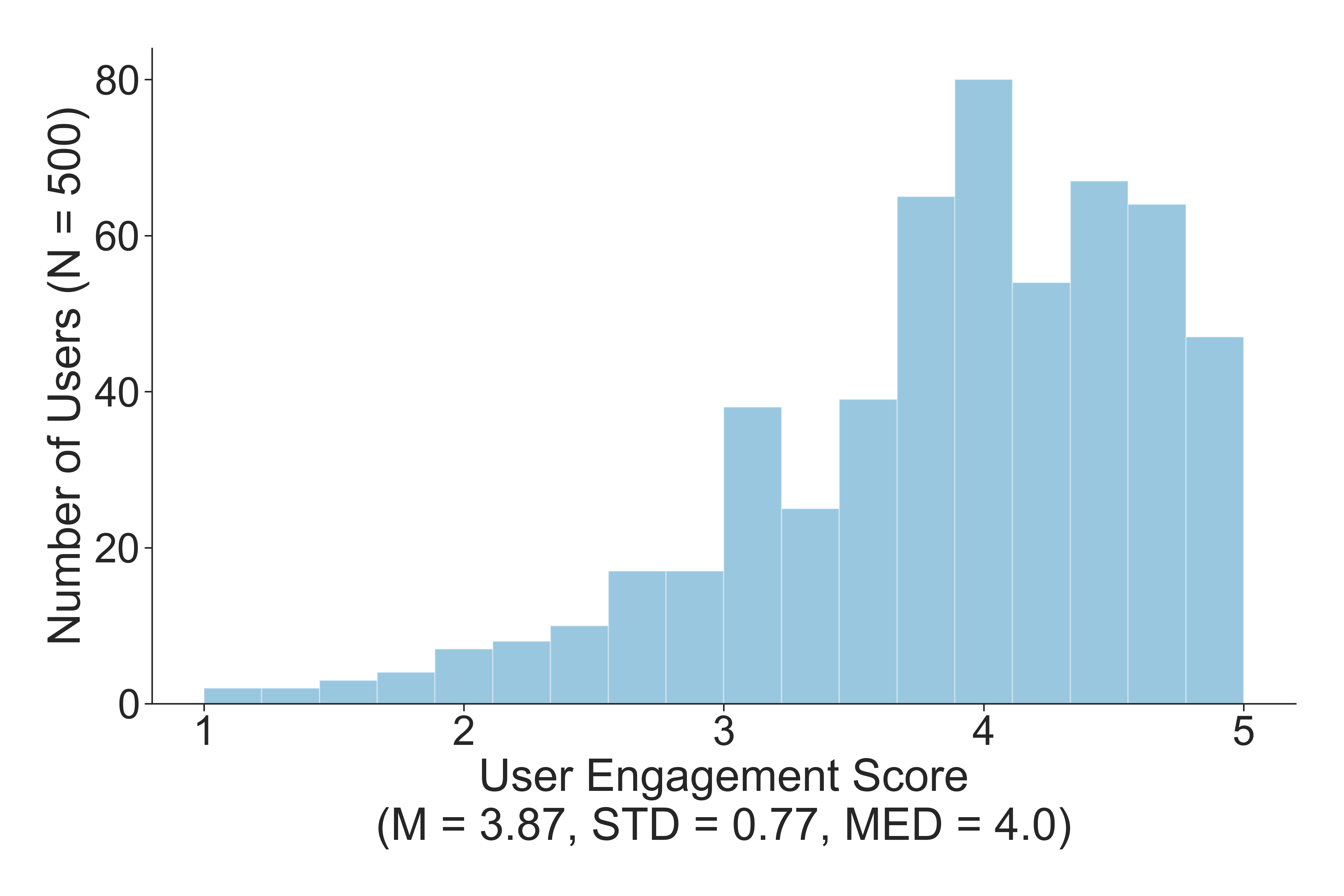}
        \caption{Distribution of engagement scores reported by audience members}
        \Description{The plot shows a right-skewed distribution of engagement scores.}
        \label{fig:ues_distribution}
    \end{subfigure}
    \hfill
    \begin{subfigure}[b]{0.49\textwidth}
        \centering
        \includegraphics[width=\textwidth]{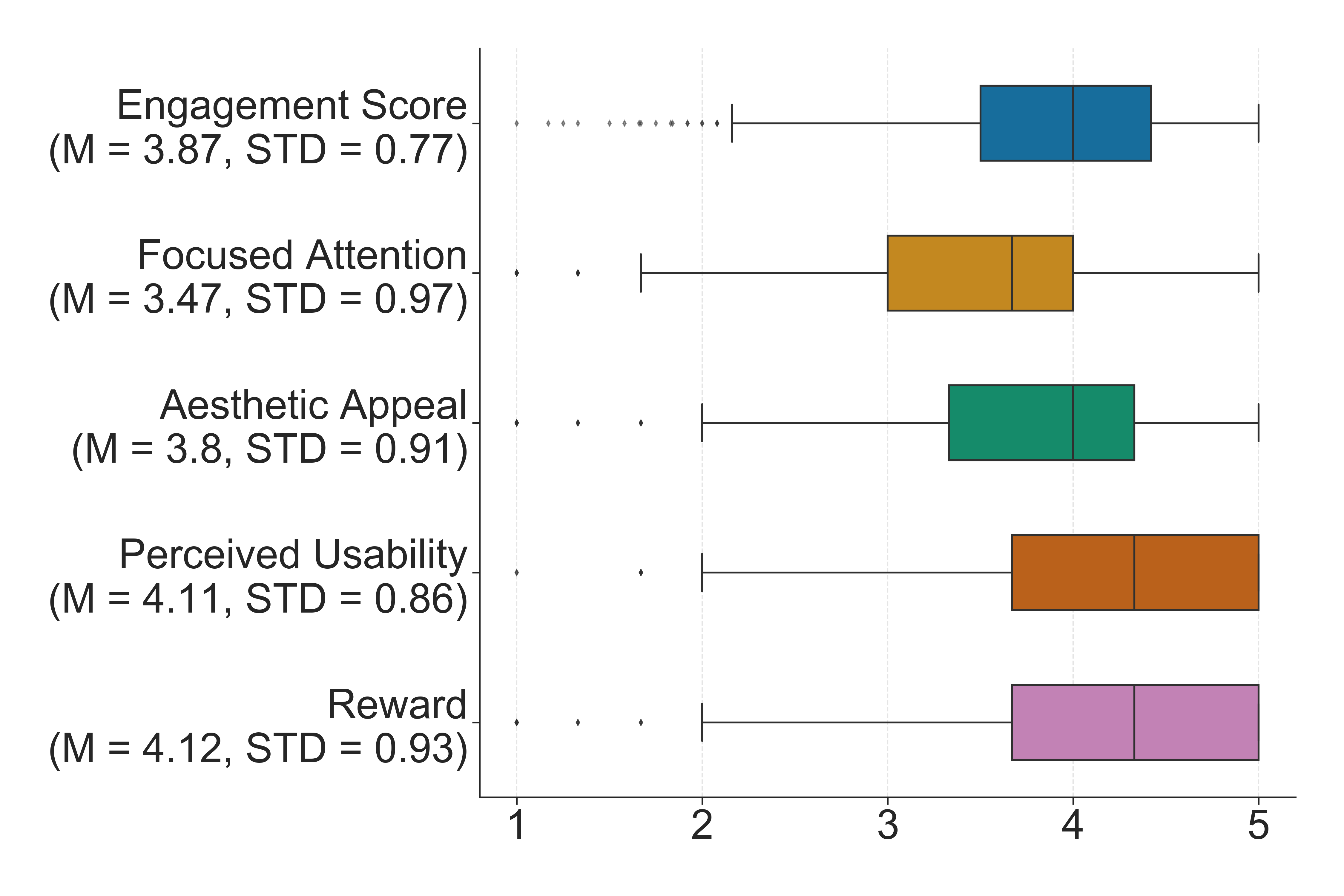}
        \caption{Distributions of the individual UES factors}
        \Description{The plot shows the distributions of the four factors from the UES survey and 
        shows both perceived usability and reward being right-skewed, while focused attention and 
        aesthetic appeal are more normally distributed.}
        \label{fig:ues_factors}
    \end{subfigure}
    \caption{Distributions of engagement scores \& UES factors}
    \label{fig:three graphs}
\end{figure}

In total, 500 members of the audience chose to take part in the study by submitting a response to the UES survey. The distribution of their engagement scores is shown in Figure \ref{fig:ues_distribution}, and the scores on the four engagement factors in Figure \ref{fig:ues_factors}. Overall, the audience were highly engaged with the experience ($M = 3.87, STD = 0.77$). However, due to the survey placement - post-credits - the sample captured could be skewed, and less engaged users are less likely to reach this part of the show. Nevertheless, there is still a proportion of users that were not engaged by the experience (recording scores below three, 14.77\%, $M = 2.48, STD = 0.49$), suggesting that even if a user reaches the end of the experience, it does not necessarily mean they were engaged - we hypothesise reasoning for this later. On the whole, as shown in Figure \ref{fig:ues_factors}, the audience found the experience to be usable ($M = 4.10, STD = 0.85$) and they felt reward in using it ($M = 4.12, STD = 0.93$). However, the audience felt that the experience did not capture their attention to a high level ($M = 3.46, STD = 0.97$), a crucial aspect of media creation. Similarly, the audience varied their opinion on the aesthetics of the experience ($M = 3.79, SD = 0.91$). From the 500 participants, 310,800 ($M = 621.60, STD = 379.83$) interaction events were recorded.

\subsection{Relationships to Engagement}
We found minor positive and negative correlations between the majority of metrics and engagement scores. Session length had the strongest correlation with the overall engagement scores ($r_s = .32, p < .001$), and time to completion was second strongest ($r_s = .30, p < .001$), both indicating a relationship between the time spent on the experience and engagement. These two metrics are related but differ slightly as the session length is how long the user`s session lasts, from start to finish, whilst the time to completion metric is the time it takes for the user to reach a defined endpoint.

\begin{figure}[h]
    \centering
    \includegraphics[width=\linewidth]{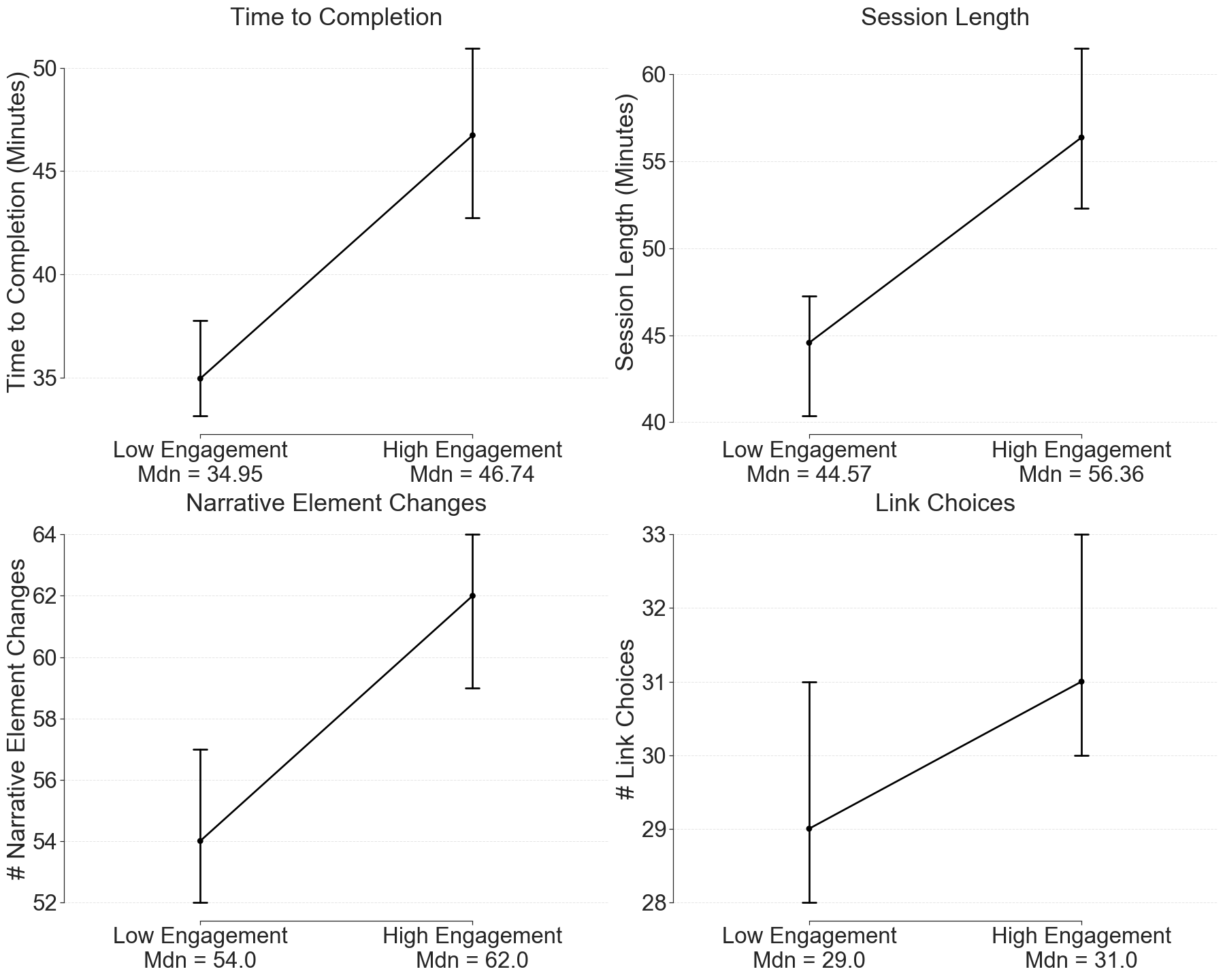}
    \caption{Estimations of the central tendency for the four metrics that are significantly 
    different between the two engagement groups. The median was used as the estimator.}
    \Description{}
    \label{fig:point_plot}
\end{figure}

When applying Mann-Whitney`s $U$ test, with correction for multiple tests, to each metric between the two engagement groups ($N_{low} = 259, N_{high} = 241, low \leq 4.0 < high$), we found a significant difference between low and high engagement with respect to time to completion ($U(N_{low} = 259, N_{high} = 241) = 22443.00, p < .001, f = .64$) and session length ($U(N_{low} = 259, N_{high} = 241) = 22111.00, p < .001, f = .65)$). For metrics relating to interaction events, we found that the number of narrative element changes, recorded when the user moves from one story element to another, are statistically different: $U(N_{low} = 259, N_{high} = 241) = 23527.00, p < .001, f = .61$. The distribution of link choices, recorded when a user actively chooses a presented option, were also significantly different between engagement levels: $U(N_{low} = 259, N_{high} = 241) = 27724.50, p < .05, f = .54$. These differences between the two groups are shown in Figure \ref{fig:point_plot}, which demonstrates that users in the high engagement group typically have longer sessions, take more time to complete, record more narrative element changes, and make more active choices in their viewing path.

To test the ability of interaction metrics to discriminate between low and high engagement, we used the chi-square test to evaluate the probability of the observed result given the null hypothesis being true ({\bf H0}). Table \ref{tab:feature_stats} presents the statistically significant results, ordered by discriminatory power ($\chi^2$). We see that the number of next buttons (the function to skip through narrative elements) is the most discriminatory metric, suggesting that users in the low engagement group skip through content much more frequently than those in the high engagement group (indicated by the direction of the relationship). Further, the second-ranked metric, short pauses, demonstrates a significant difference in the way that users consume the media. Users in the low engagement group were more likely to record shorter periods between interactions; in contrast, users in the high engagement group were more likely to record very long pauses - indicating they likely spend more time watching content. 

\begin{table*}[h!]
    \caption{Discriminatory statistical features (significance level: ***= $p < .001$, *= $p < .05$). 
    $pr_l$ and $pr_h$ denote the Pearson residuals for the feature for low and high engagement. 
    $total_l$ and $total_h$ represent the observed frequency for each feature in the low and high 
    engagement classes. Direction represents how the relationship is weighted, a negative direction 
    means the low engagement group recorded more of the feature.}
    \label{tab:feature_stats}
        \begin{tabular}{@{}ccccccc@{}}
            \toprule
            \textbf{Feature}          & \textbf{$\chi^2$} & \textbf{$pr_l$} & \textbf{$pr_h$} & \textbf{$total_l$} & \textbf{$total_h$} & \textbf{Direction} \\ \midrule
            Next Button               & 175.09***        & 9.23              & -9.48              & 2388                 & 1464                  & -924                          \\
            Short Pauses              & 84.06***         & 6.39              & -6.56              & 1230                 & 767                   & -463                          \\
            Narrative Element Changes & 39.67***         & -4.39             & 4.51               & 14626                & 14918                 & 292                           \\
            Medium Pauses             & 36.33***         & 4.20              & -4.31              & 731                  & 488                   & -243                          \\
            Long Pauses               & 24.38***         & 3.44              & -3.53              & 566                  & 388                   & -178                          \\
            Play/Pause                & 11.59***         & -2.37             & 2.43               & 742                  & 835                   & 93                            \\
            Very Long Pauses          & 5.57*            & -1.64             & 1.69               & 1245                 & 1296                  & 51                            \\
            Fullscreen                & 4.35*            & -1.45             & 1.49               & 314                  & 350                   & 36                            \\ \bottomrule
        \end{tabular}
\end{table*}

\subsection{Predicting Engagement}
\begin{figure}[h]
    \centering 
    \includegraphics[width=\linewidth]{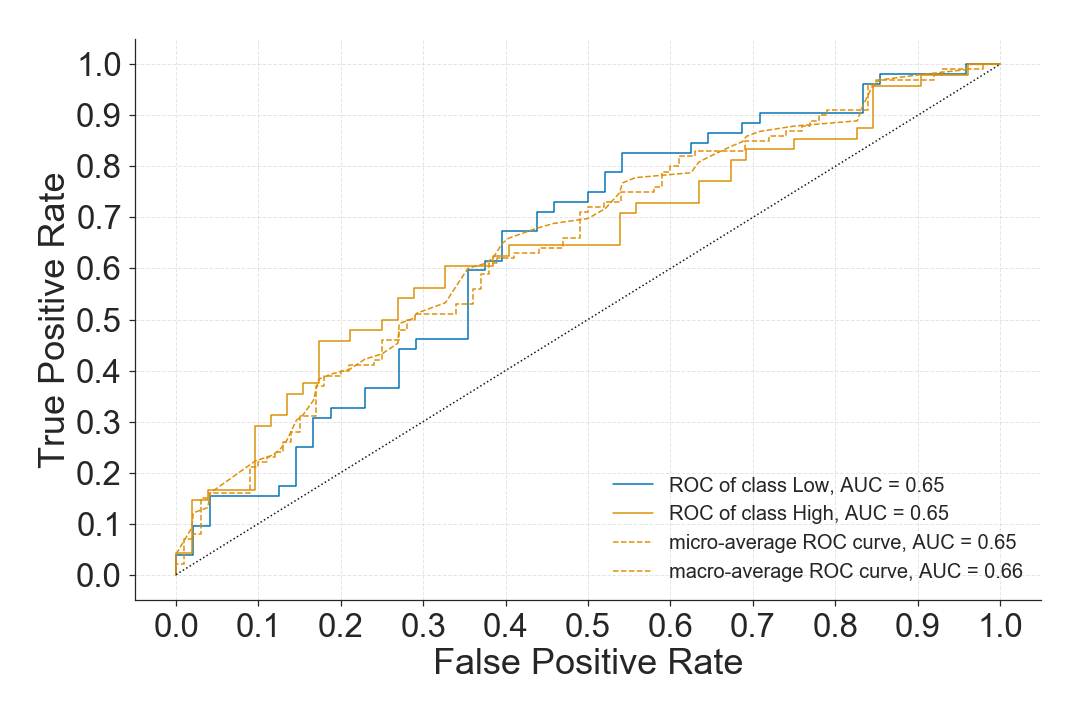}
    \caption{Receiver Operating Characteristic (ROC) curve demonstrating the model's ability 
    to distinguish between low and high engagement}
    \Description{ROC plot demonstrating the model performance, with a macro-average of 0.66}
    \label{fig:roc_auc}
\end{figure}

When evaluating which model performed best in predicting engagement levels, we found that Logistic Regression (LR) performed the best across all evaluations (LR $\mu AUC = .60$, all other models: $\mu AUC = 0.55-0.60$), and as it is a comparatively simpler and more interpretable model, we chose it to be the model that we used for the rest of the analysis. To tune the model to our data, we evaluated the best hyperparameter values for the normalised penalty ($L_1$ and $L_2$), the strength of regularisation, and the stopping criteria tolerance. We found that the most optimal ($AUC = .61$) hyperparameter configuration was: $L_2$ normalisation penalty, a regularisation strength of 1, and a stopping tolerance of $1 \times 10^{-5}$. Fitting an optimised model on a training set consisting of 400 samples and testing on 100 samples, we observed that the model can separate between the two engagement classes ($AUC = .66$, \emph{precision} $= .61$, \emph{recall} $= .61$, \emph{f1-score} $= .61$), as shown in Figure \ref{fig:roc_auc}. We computed a paired $t$-test to compare the performance of the model with a baseline that generates predictions uniformly randomly, finding that the Logistic Regression model performs significantly better ($t(32) = 4.47, p < .01$). These results suggest that engagement can be modelled and accurately predicted using interaction data collected from an interactive media experience. 

\begin{figure}[h]
    \centering
    \includegraphics[width=\linewidth]{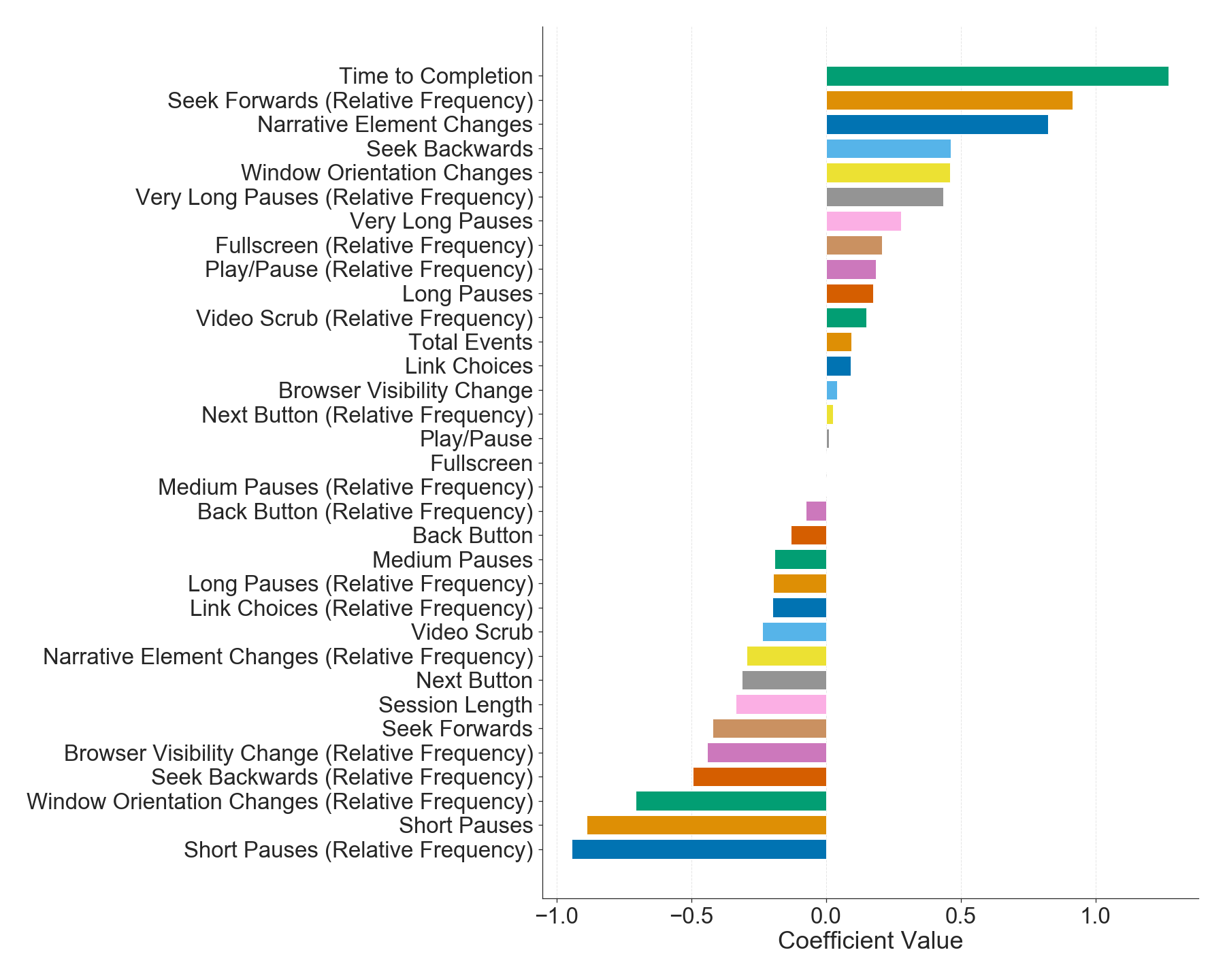}
    \caption{Feature coefficients when predicting engagement.}
    \Description{A plot showing the model coefficients, where time to completion is the strongest
    positive coefficient and the relative frequency of short pauses is the strongest negative 
    coefficient.}
    \label{fig:feature_coeff}
\end{figure}

To assess metric importance, we ranked each using regression coefficients, which describe size/direction of the relationship between a metric and target (Figure \ref{fig:feature_coeff}). Coefficients with a large positive value increase probability of predicting high engagement; large negative values increase probability of predicting low engagement. 

We observed (Figure \ref{fig:feature_coeff}) that an increase in the time to completion metric correlates to higher engagement, supporting the significant difference between high and low engagement seen in Figure \ref{fig:point_plot}. Similarly, an increase in narrative element changes is positively associated with higher engagement; however, as the relative frequency of narrative element changes increases, the prediction output weights towards low engagement. The importance of very long pauses and short pauses (both counts and relative frequencies) is demonstrated by high engagement being positively associated with a large number of very long pauses, and negatively associated with  a large number of short pauses. In contrast, a longer time to completion, higher count of narrative element changes, and more very long pauses are associated with high engagement prediction. The importance of temporal metrics is demonstrated; the time between events, rather than the events themselves, is important. We also observed that the occurrence of a window orientation change event is associated with higher engagement, but a high frequency of such events is associated with low engagement.

\begin{figure}[h]
    \centering
    \includegraphics[width=\linewidth]{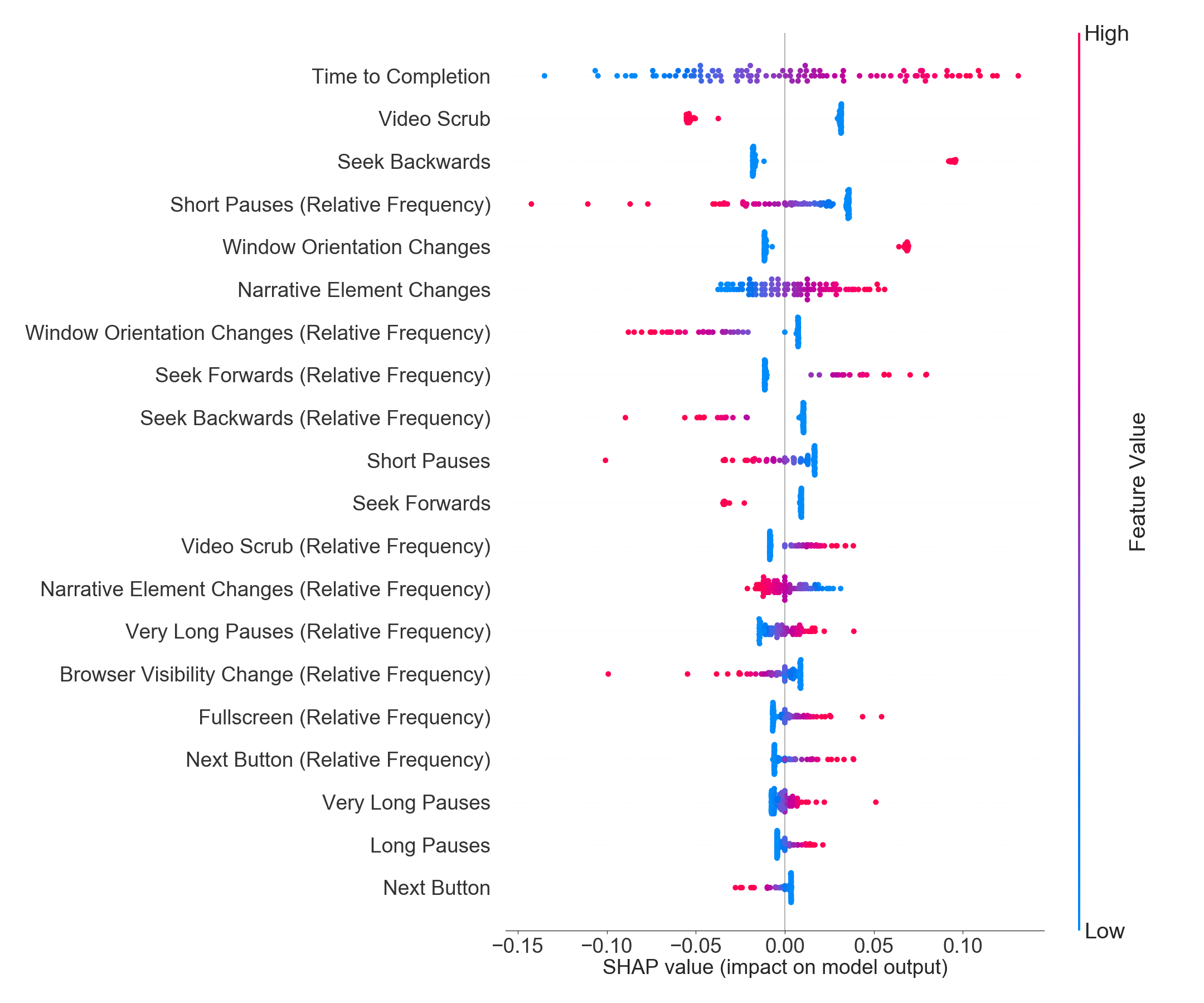}
    \caption{Feature importance based on contribution to model output
    (SHAP values). The position of the metrics on the $y$-axis is ordered by the sum of SHAP values across all samples.}
    \Description{A plot showing the model coefficients, where time to completion is the strongest
    positive and the relative frequency of short pauses is the strongest negative 
    coefficient.}
    \label{fig:shap}
\end{figure}

To further explore the interaction metrics contributions to the prediction of high and low engagement, we calculated SHAP values -- results are shown in Figure \ref{fig:shap}. The results reveal that a high value (indicated by the red colour) for the time to completion metric increases the probability of predicting high engagement, while a lower value increases the probability of predicting low engagement, further corroborating the findings shown in Figure \ref{fig:point_plot}. The second most important feature, video scrubs, shows that there are two groups of users: the first record higher values which result in pushing the prediction towards low engagement, whilst the second record smaller values which push the prediction towards higher engagement. Similarly, an increase in seeking backwards (rewinding the video by ten seconds) events results in higher engagement, while smaller values have little contribution to the model output.

This work gives insight into the effect of the interaction metrics on predicting engagement; crucially, it highlights the importance of temporal metrics in understanding users' engagement: the amount of time it takes a user to complete is indicative of their engagement. The types of pauses between user actions also help to show how users consume the media: short pauses associate with low engagement, very long pauses with high engagement. More generally, the usefulness of the metrics to garner insights on user engagement with an interactive media experience is positively demonstrated.
\section{Discussion}\label{sec:discussion}
This paper presents a study on interaction data and engagement metrics collected from a production-quality, nationally released interactive TV show to answer the question: \emph{what behaviours can we infer from interaction metrics that help in understanding how engaged an audience member is with an interactive media experience?}. There is a lack of existing knowledge about techniques to measure engagement in this new form of media; our objective is to develop a rich understanding of engagement, user behaviours and how they relate. Ultimately, this knowledge should aid media creators to create better experiences, to understand the audiences attracted to new forms of content, and to improve user experience. 

We found that a higher next button count, narrative element change frequency, and shorter pauses between interactions were related to low engagement. The combination of these metrics may indicate a skipping-type behaviour, where the user moves through the content without watching in depth. For content creators, it is important to identify loss of engagement during the experience and detecting skipping in real-time could provide an opportunity to modify an experience on-the-fly to provide a shorter, summarised version of the narrative. A higher rate of skipping through content provides insight about the performance of experiences - if users skip through without watching the content, then it may indicate a problem with or a dislike of the content. However, whether the user is skipping because they are not engaged with the content or whether they are unable to engage with the content because they are skipping is unclear and requires further investigation.

We also found that whilst a single window orientation change event was positively associated with high engagement, a high relative frequency of such events is associated with low engagement. We hypothesise that a single orientation change - e.g., switching to horizontal - shows settling-in behaviour, whilst switching repeatedly demonstrates an indecisiveness or an unsatisfactory experience on the device being used. This type of behaviour could enable creators to actively monitor and react, collecting more data from the individual to understand whether there is an issue with the content or to signal that the story is best experienced in a particular orientation to improve overall user experience.

The best predictor of engagement was found to be time to completion, where lower values relate to lower engagement and higher values to higher engagement. Using time to completion in combination with other metrics suggest that high engagement is associated with a consumption-type behaviour (longer time to completion, higher count of narrative element changes, and more very long pauses). Monitoring for increases in these metrics could provide content creators with a detailed picture of the success of their experience and the means to retrospectively understand what types of story experiences work better.

The work has several limitations. Drawing actionable conclusions from the findings is constrained by small sample size, limiting generalisability; we plan to address this by collecting additional data and testing the findings on other interactive media experiences. While there were tangible benefits to collecting data from a live piece of content rather than a controlled environment, it gives limited knowledge about the audience, e.g., whether audience members visited Click or completed the survey more than once; we have developed a technique to avoid this issue in future studies. There is also a question as to how representative the interactions of the low engagement users are of the wider un-engaged audience, as only a particular type of less engaged user will reach the end and complete the survey. The study and results presented in this paper focused on data collected from a single experience, and we have not evaluated whether these results would generalise to others - we plan to test this in future work. Finally, whilst we identified and inferred a range of behaviours, we did not validate them; we plan to address this in future observational studies, along with collecting additional data on user intent before using an interactive media experience.

\section{Conclusion}\label{sec:conclusion}
Media creation is moving from a traditional one-size-fits-all narrative to a more personalised experience which can adapt to audience context, knowledge, and needs. As movie pioneers had to discover and develop the craft of film-making, creators of interactive content must learn what works and what does not in this new domain. To do so, it is essential to better understand how different aspects of their production -- the interaction design, the ways in which it adapts, and the content itself -- impact on audience experience, of which a key measure will be engagement. 

Measuring engagement is challenging, but an approach from other domains that is cheap and works at scale is capturing user interactions and infer engagement from these. To test whether this approach can be applied in this domain, we undertook a study using a nationally released, production-quality interactive TV show. We found temporal metrics - the time between events and the time it takes an audience member to complete - are both predictive of, and can effectively distinguish, high and low engagement. More specifically, media creators can gain an understanding of low engagement by monitoring the relative frequency of next button usage and narrative element changes along with the number of short pauses between interactions. The amount of time an audience member takes to complete the experience, the number of narrative element changes, and the presence of very long pauses between interactions appear to be indicative of high engagement. 

Our findings demonstrate the importance of considering the time between events, and not just the events themselves, and the significance of temporal metrics when attempting to understand engagement with interactive media experiences. Overall, we highlight points of consideration for media creators, metrics that are useful to monitor and collect to understand levels of engagement with new experiences, and identified behaviours that link interaction metrics to engagement.

\begin{acks}
Jonathan Carlton would like thank the EPSRC and the BBC for iCASE PhD funding support (16000156).
\end{acks}

\balance

\bibliographystyle{ACM-Reference-Format}
\bibliography{references}

\end{document}